\documentclass[journal]{IEEEtran}
\usepackage{times}
\usepackage{amsmath,epsfig}
\usepackage{amssymb}
\usepackage{amsfonts}
\usepackage{array,dcolumn}
\usepackage[ruled,vlined,linesnumbered]{algorithm2e}
\usepackage{subcaption}
\usepackage{psfrag}
\usepackage{amsmath}
\usepackage{amsfonts}
\usepackage{amssymb}
\usepackage{balance}
\usepackage{accents}
\usepackage{booktabs, adjustbox}
\usepackage{cite}
\usepackage[all]{xy}
\usepackage{dsfont}
\usepackage{hyperref}
\usepackage{url}
\usepackage{xcolor}
\usepackage[nolist]{acronym}
\usepackage{graphicx}
\usepackage{float}
\usepackage{threeparttable}
\usepackage{cancel}
\usepackage[official]{eurosym}
\usepackage{chemformula}
\usepackage{subcaption}
\usepackage{graphicx}
 
\newcommand{\ilm}{\textcolor{black}}

\newcommand{\PP}[1]{\textcolor{black}{#1}}
\newcommand{\ppp}[1]{\textcolor{black}{#1}}
\newcommand{\LD}[1]{\textcolor{black}{#1}}
\newcommand{\AL}[1]{\textcolor{black}{#1}}
\newcommand{\rv}[1]{\textcolor{black}{#1}}
\newcommand{\dd}{\mathop{}\!\mathrm{d}}
\include{pythonlisting}
\def\BibTeX{{\rm B\kern-.05em{\sc i\kern-.025em b}\kern-.08em
    T\kern-.1667em\lower.7ex\hbox{E}\kern-.125emX}}
    
\title{\rv{Modeling and Analysis of Data Trading on Blockchain-based Market in IoT Networks}}

\author{Lam Duc Nguyen, \textit{Student Member, IEEE}, Israel Leyva-Mayorga, \textit{Member, IEEE}, \\ Amari N. Lewis, \textit{Student Member, IEEE}\thanks{\rv{This work has been in part supported by the European Research Council (Horizon 2020 ERC Consolidator) under Grant 648382 WILLOW, by the European Union's Horizon 2020 program under Grant 957218 IntellIoT, the Independent Research Fund Denmark (DFF) under Grants Nr. 8022-00284B (SEMIOTIC)
and Nr. 9165-00001B (GROW), and the National Science Foundation Graduate Research Fellowship under Grant DGE-1839285.}}\thanks{Lam Duc Nguyen, Israel Leyva-Mayorga, and Petar Popovski are with the Connectivity Section, Department of Electronic System, Aalborg University, Denmark. Contact: \{ndl, ilm, petarp\}@es.aau.dk \textit{(Corresponding Author: Lam Duc Nguyen)}}\thanks{Amari N. Lewis is with the Donald Bren School of Information and Computer Sciences, University of California Irvine, USA. Contact: amaril@ics.uci.edu  }, and Petar Popovski, \textit{Fellow, IEEE}}%

\begin{document}
\maketitle

\begin{abstract} 
Mobile devices with embedded sensors for data collection and environmental sensing create a basis for a cost-effective approach for data trading. For example, these data can be related to pollution and gas emissions, which can be used to check the compliance with national and international regulations. The current approach for  IoT data trading relies on a centralized third-party entity to negotiate between data consumers and data providers, which is inefficient and insecure on a large scale. In comparison, a decentralized approach based on distributed ledger technologies (DLT) enables data trading while ensuring trust, security, and privacy. However, due to the lack of understanding \rv{of} the communication efficiency between sellers and buyers, there is still a significant gap in benchmarking the data trading protocols in IoT environments. Motivated by this knowledge gap, we introduce a model for DLT-based IoT data trading over the Narrowband Internet of Things (NB-IoT) system, intended to support massive environmental sensing. We characterize the communication efficiency of three basic DLT-based IoT data trading protocols via NB- IoT connectivity in terms of latency and energy consumption. The model and analyses of these protocols provide a benchmark for IoT data trading applications.
\end{abstract}


\section{Introduction}
In 2025, the volume of sensing data generated by personal IoT devices is expected to reach 79.4 ZB globally\cite{data794}. Many attempts have been made to improve and adapt business workflows to exploit the availability of IoT data\cite{previous1,previous2}; among these, IoT data trading is the most popular approach. \PP{Various services for trading of IoT data are emerging, connecting various devices and distributed IoT data sources, thereby facilitating data providers to exchange their data \cite{infocom2019}.} 

\PP{Interesting use cases for data trading include public transport systems, for example, the bus network in Aalborg, Denmark.} In these \AL{systems}, the density of personal travel card swipes at specific bus stations could be useful information, not only to the administration of transport systems, but also to the local taxi companies. The taxi companies benefit from the data of anomalous passenger traffic patterns for the purposes of improving ride-sharing and private services\cite{mit}. Also, analyzed traffic data of passengers can be collected via IoT infrastructure and recommendation services to taxi companies can be sold. Besides,  drivers can exchange information about the traffic status of \AL{a} particular street with others to avoid traffic jams or \AL{to} exchange \AL{green house} gas emission information with manufacturers. Hence, IoT data can be considered as a tradable digital asset.

\PP{Traditional trading systems (e.g. Paypal) feature a single point of failure, the lack of trust, transparency, and incentive for data trading, which is preventing the availability of digital information from data providers to customers.} On the other hand,   \PP{Distributed ledger technologies (DLTs) and Blockchains\footnote{The terms DLT and \emph{Blockchain} will be used interchangeably throughout this paper, Blockchains are a type of DLT, where chains of blocks are made up of digital pieces of information called transactions and every node maintains a copy of the ledger} support immutable and transparent information sharing among involved untrusted parties \cite{bitcoin}. Outside of its role in financial transactions, DLTs are seen }as a key enabler for trusted and reliable distributed monitoring systems. The authentication process for DLTs relies on consensus among multiple nodes in the network\cite{iotbc}. In Blockchain-enabled IoT networks \cite{witness}, transactions can include sensing data, or monitoring control messages, and these are recorded and synchronized in a distributed manner in all the participants of the system. These participants are called miners or peers and, in some specific DLTs, users are charged a transaction fee to deploy and execute transactions. 

In addition, DLTs allow the storage of all transactions into immutable records and every record is distributed across many participants. Thus, security in DLTs comes from the decentralized operation, but also {from} the use of strong public-key cryptography and cryptographic hashes. The benefits of the integration of DLTs into IoT data trading systems include: i) guarantee of immutability and transparency for environmental sensing data, ii) removal of the need for third parties, iii) development of a transparent system for heterogeneous IoT data trading networks to prevent tampering and injection of fake data from the stakeholders\cite{iotmagazine}.

With the spread of ubiquitous marketplaces, \PP{it became relevant to explore} the use of IoT data trading in marketplace environments. For instance, in \cite{IoTMartetplace}, Gupta et al. introduced the architecture for a dynamic decentralized marketplace for trading IoT data. The approach involves a 3-tier design: 1) provider, 2) broker and 3) consumer. The use of DLTs in their work is primarily to manage the terms of agreement between involved parties. Additionally, a reputation system is used in the design to penalize the participants and reduce their rating. Bajoudah et al. present a marketplace model and architecture for the  trading of IoT streaming data in \cite{IoTMartetplace2}. Within their work, periodic checkpoints during data exchange are introduced to limit fraudulent activity on either side. In \cite{Missier}, Missier et al. propose another marketplace, where streams of IoT data are the main assets traded utilizing Oracles for the off-chain queries. Xiong et al. \cite{Xiong} present a
\PP{trading mode based on smart contracts. It incorporates machine learning to guarantee fairness of data exchange} and utilizes arbitration institution to deal with the dispute over the data availability in the data trading. However, the arbitration institution in the trading mode is a trusted entity of trading parties.
\rv{Dai et al. \cite{Dai} introduced a secure data trading ecosystem based on Blockchain by combining the Intel Software Guard Extensions (SGX). The proposed ecosystem securely processes the data, but, the data source and analysis results highly depend on a trusted SGX-based execution environment.}
%
\rv{In \cite{feng2018competitive}, the authors proposed a decentralized Blockchain-based platform for data storage and trading in a wireless powered IoT crowd-sensing system. The data from RF-energy beacons are transmitted to the ledger for decentralized services, which supports the analytical condition for valuable results about the equilibrium strategies in the distributed systems.}

The related work indicates a \rv{knowledge} gap in terms of: 1) a benchmark for IoT data trading, and 2) analysis of the cost of IoT data trading in terms of communication, specifically in city-level networks. 
The efficiency of a Blockchain-based data trading protocol is a major concern for data traders. 
\rv{Future markets will be highly dynamic and low latency trading is critical} \rv{to maximize} the efficiency of the marketplace. However, currently there is a lack of a general framework that provides a guideline for the use of trading protocols based on a set of neutral and commonly accepted rules. \PP{A proper benchmark helps the interested parties to understand the tradeoffs in Blockchain-based systems and the associated performance indicators.}

In this paper, first, we design a DLT-based trading system for exchanging IoT data. \PP{We have chosen the NB-IoT standard~\cite{nbiotsmartcity} as the underlying connectivity solution, as it is seen} by the mobile operators as a major candidate to dominate wide-range connectivity for future smart cities. 
\rv{Unlike \ppp{many} other IoT technologies, NB-IoT is able to offer symmetric uplink/downlink throughput, which is an essential feature from the viewpoint of a DLT \cite{3GPPTS36.213,iotbc}}.
The proposed trading system includes the following IoT data trading protocols; \textit{General Trading (GT)}, \textit{Buying on Demand (BoD)}, and \textit{Selling on Demand (SoD)}.  Here\rv{,} we use the term ``on demand'' from the \rv{perspective} of the smart contracts that implement the transactions between buyers and sellers.  Each trading protocol is customized for different scenarios. \PP{\textit{GT} could be considered as the usual trading protocol in \rv{the} data marketplace, while the \textit{BoD} and \textit{SoD} are protocols used to support particular demands from either sellers or buyers.}

\PP{The analysis and simulation results show that the \textit{GT} protocol has outstanding performance in terms of latency and energy consumption; however, it requires mechanisms to guarantee the continuous availability of data. 
On the contrary, the \textit{BoD} protocol can be \ilm{implemented in} Vehicle-to-Infrastructure (V2I) networks, where vehicles can trade their emission information with manufactures. \rv{ Finally, the \textit{SoD} protocol is particularly useful when customers are interested in collecting specific data, \ppp{which, however, may not be immediately }available on the market}. \rv{This protocol} can also be deployed in Vehicle-to-Vehicle (V2V) networks where the drivers want to buy traffic jam information of a specific street from other vehicles on the road. Clearly, \textit{SoD} protocols, on their own, would face situations in which the data is no longer available for customers after the initial advertising phase.
In practice, the three trading protocols present interesting synergies and can be implemented together in a single system, which will select the best one based on the actual situation.} 

\PP{The contributions of this paper can be stated as follows. First,  we present 
a solution for \rv{a} systematic DLT-based IoT data smart trading towards city-level networks using NB-IoT connectivity. Next, we propose three IoT data trading protocols namely \textit{General Trading (GT)}, \textit{Buying on Demand (BoD)}, and \textit{Selling on Demand (SoD)}. The cost model of each trading protocol is derived and analyzed along with NB-IoT connectivity. Both resources consumed by executing DLT/smart contracts and NB-IoT devices are investigated. Finally, the analysis and the associated experimental results provide a benchmark for data trading protocols in wide-area IoT networks.} 

The remainder of this paper is organized as follows. In the next section, we outline the general architecture of DLT-based trading system and introduce three IoT data trading protocols. Then, we present the system model, \PP{including the physical deployment of the devices}. In Section III and IV, we model and analyze the performance of Blockchain-enabled IoT network in terms of latency and energy. Then, we evaluate and prove the derived model and design in Section V. Finally, we conclude the paper in Section VI. 

\section{DLT-enabled IoT Data Smart Trading Architecture and Protocols}
\PP{This section presents the general system model of DLT-based IoT data trading as well as the data trading system with the three protocols tailored to different scenarios. Table\,\ref{tab:symbols} summarizes the used notation.}

\begin{figure}[t]
    \centering
    \includegraphics[width=0.9\linewidth]{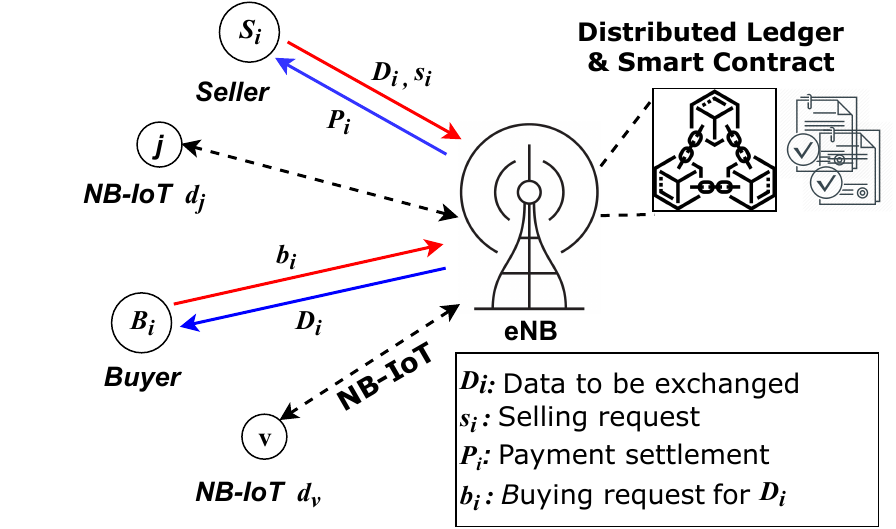}
    \caption{General system model of DLT-enabled IoT Data Trading via NB-IoT connectivity where seller $\mathcal{S}_i$ and buyer $\mathcal{B}_i$ make a deal on the data $D_i$. }
    \label{fig:fig1}
\end{figure}

\subsection{DLT-enabled Data Trading via NB-IoT}

The general architecture of DLT-based IoT data trading includes three main components: data providers (sellers  $\mathcal{S}$), data consumers (buyers $\mathcal{B}$) and a distributed ledger, shown in Fig.\,\ref{fig:fig1}. Each seller or buyer can own one or more devices in the network. Here we assume that buyers and sellers act as digital wallets in a distributed network. \PP{During a trade  denoted by $\mathcal{T}_i$,} the seller $\mathcal{S}_i\in\mathcal{S}$ and buyer $\mathcal{B}_i\in\mathcal{B}$ communicate using the wide-area NB-IoT links. The trading procedure occurs to complete a deal between $\mathcal{S}_i$ and $\mathcal{B}_i$,  exchanging data $\mathcal{D}_i\in\mathcal{D}$ and payment $\mathcal{P}_i$. First, $\mathcal{B}_i$ completes the payment $\mathcal{P}_i$ to \PP{$\mathcal{S}_i$} in reference to the requested data, $\mathcal{D}_i$, and \PP{$\mathcal{S}_i$ delivers $\mathcal{D}_i$ to $\mathcal{B}_i$ immediately}. The general procedure from Fig.\ref{fig:fig1} can be described as follows:

\subsubsection{Buyer $\mathcal{B}_i$} Subscribes to the IoT data in distributed ledger generated and published by $\mathcal{S}_i$, and $\mathcal{B}_i$ makes a data request, $b_i$ regarding its preferred data, $\mathcal{D}_i$. The $b_i$ will be transmitted to $\mathcal{S}_i$ and recorded in the ledger via transaction $T_{i, add}$ 
for negotiation based on factors such as amount of data, quality of data, price, discount, etc. After choosing \PP{$\mathcal{D}_i$} from the list, $\mathcal{B}_i$ generates a transaction $T_{i, commit}$ which executes payment from $B_i$'s wallet. Once \textbf{$\mathcal{B}_i$} receives the \PP{$\mathcal{D}_i$} via $T_{i, settle}$, it will generate a confirmation back to ledger. 

\subsubsection{Seller $\mathcal{S}_i$} 
Has two main roles; to collect data from the environment (e.g., environmental sensing data, geographical data or data from surveillance systems) and to act as a hub gathering data from neighboring devices to sell on the market.
$\mathcal{S}_i$ aims to earn the payment $\mathcal{P}_i$ from $\mathcal{B}_i$ by delivering $\mathcal{D}_i$ to $\mathcal{B}_i$.  After publishing a hashed version of its data and prices to the market via $T_{i, add}$, $\mathcal{S}_i$ waits for buying requests. Based on the predefined rules in the smart contract system, upon receiving a request from $\mathcal{B}_i$ and the appearance of $T_{i, commit}$, generated by $\mathcal{B}_i$, 
the seller $\mathcal{S}_i$ can receive the payment $\mathcal{P}_i$. Finally, it confirms to the ledger that the trade $\mathcal{T}_i$ is \rv{complete.} 

\begin{table}[t!]
\centering
\caption{NOMENCLATURE}
 \begin{tabular}{@{}  p{1.4cm}  p{4.8cm}  p{1.5cm}  @{}}
 \toprule
 \textbf{Parameters} & \textbf{Descriptions}  & \textbf{Values} \\ [0.5ex] 
 \midrule
 \textbf{General} &  & \\
$N$ & Total number of NB-IoT devices & 10000 \\
$M$ & Number of DLT miners & $\leq 20$ \\
$\mathcal{S}$ & Set of Data Providers (Sellers) & $|\mathcal{S}|\leq 10^4$\\
$\mathcal{B}$ & Set of Data Consumers (Buyers) &  $|\mathcal{B}|\leq 10^4$\\
$\mathcal{D}$ & Data to buy or sell & -- \\
$\mathcal{T}=\{\mathcal{T}_i\}$ & Set of trades & --\\
$\lambda^u$ & Uplink request arrival rate & Eq.3 \\
$\lambda^d$ & Downlink request arrival rate & Eq.3\\
$G_t, G_r $ & Transmitter and receiver antenna gains & 1 \\
$\beta$ & Path loss exponential & $\{2.4, 2.7,$\\
&&$\quad3.0, 3.3\}$ \\
$ \gamma$ & Receiver sensitivity & $3.6*10^{-10}$\\
$ \lambda_c$ & Computing speed of a miner & 0.3 \\
$ \lambda_0 $ & Scaling factor & 0.05\\
$P_c $ & Power of miner & 6 \\
$\tau$ & The unit length & 10 ms \\
$d$ & The average time interval between two NPDCCH & $\left[0.05:0.2\right]$ \\
$\mathcal{R}^u$, $\mathcal{R}^d$ & Uplink and Downlink transmission rate & - \\
 \textbf{Energy}   & & \\
$E_{\mathcal{T}_i}$ & Energy required to complete a trade $\mathcal{T}_i$ & Eq. 7 \\ 
$E^u$ & Uplink energy consumption  & Eq. 7 \\
$E^d$ & Downlink energy consumption  & Eq. 7 \\
$E_{DLT}$ & Blockchain energy consumption &  Eq. 6 \\
$E_{sync}$ &  Energy required for synchronization & -\\
$E^u_{rr}$ & Uplink: Energy for resource reservation  & Eq.11  \\
$E^u_{rr}$ & Uplink: Resource reservation energy & Eq.11 \\
$E^d_{sync}$ & Downlink: Energy for synchronization & 0.33 \\
$E^d_{rr}$ & Downlink: Resource reservation energy & Eq.11  \\
$E^d_{rx}$ & Downlink: receive energy  & Eq.17 \\
$P_l, P_I $ & Listening Power& 0.1 W\\
$P_I$ & Idle Power & 0.2 W \\
$P_t $ & Transmission Power & 0.2 W \\
$P_c$ & Power consumption in electronic circuits & 0.01 W\\
\textbf{Latency}  & & \\
$L_{W}$ & Average computation latency of a miner & Eq.19 \\
$ L_{DLT} $ & Total DLT latency  & Eq.  \\
$L_{tM}$ & DLT average transmission latency & Eq.18 \\
$L_{sync}$  & Synchronization Latency & 0.33s\\
$N_{r_{max}}$ & Maximum number of attempts &  10\\
$P_{rr}$ & Probability of resource reservation & Eq.8 \\ 
\bottomrule
\end{tabular}
\begin{tablenotes}
      \small
      \item * Eq: Equation
    \end{tablenotes}
    \label{tab:symbols}
\end{table}

\subsubsection{Distributed Ledger} The DLT manages a distributed ledger to record all data trading history which is grouped into blocks and linked together chronologically. The deployed smart contracts autonomously control the order and automate payments from parties without the need of human interaction. 
The smart contracts guarantee trust, transparency and speed of exchanging information. These can be deployed based on the negotiation between data providers and customers via $T_{i, deploy}$. Any change in smart contract\rv{s} (e.g. change of price, amount of data, or discount) can be performed via $T_{i, update}$. 

In order to minimize the cost of storage, the sensing data could be hashed and recorded at more powerful DLT nodes, and only the hash of data is recorded to ledger. Then, a message is sent back to confirm that the data has been added to the ledger. After both $\mathcal{S}_i$ and $\mathcal{B}_i$ are satisfied with the terms of the contract, $T_{i, settle}$ is executed to
get the payment $\mathcal{P}_i$ from $\mathcal{B}_i$ to transfer to $\mathcal{S}_i$'s wallet, while the data $D_i$ is transmitted to the storage address of $\mathcal{B}_i$. We assume that the data services, (e.g., data storage, trading and task dispatching) are implemented on top of a permissionless Blockchain. The sensing data are formatted into normal transactions of fixed size. To enhance efficiency, only the digest of each transaction is stored on the chain, and the content of the transactions are stored by each consensus node off-chain or at the \PP{IPFS (InterPlanetary File System) storage.}

\begin{figure*}[t]
\centering
\begin{subfigure}{0.33\textwidth}
\includegraphics[width=0.9\linewidth]{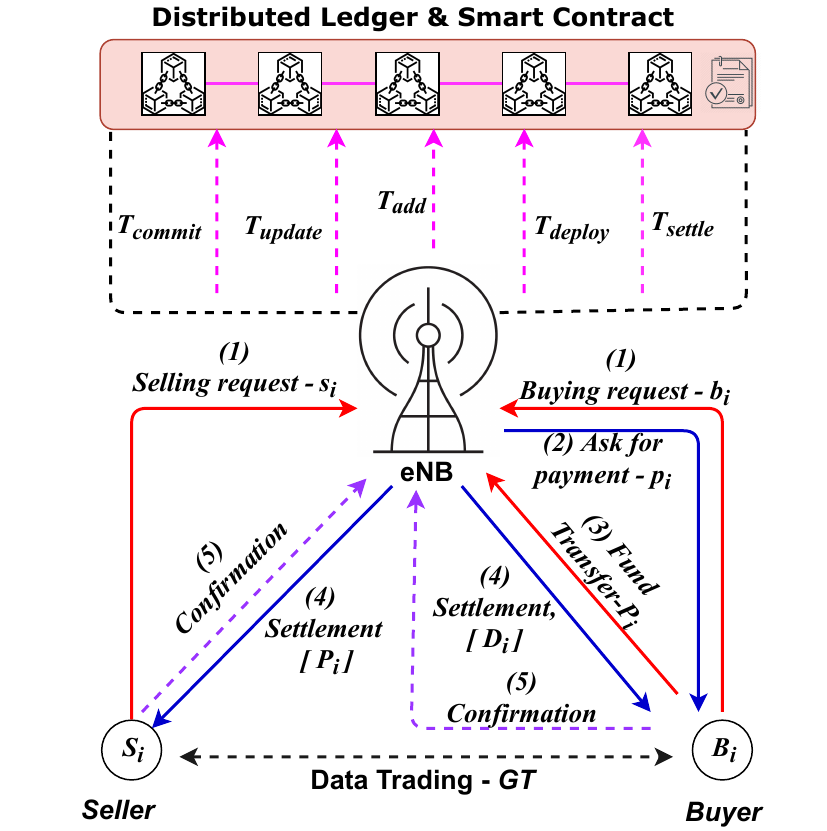} 
\caption{\textit{General Trading (GT) } }
\label{fig:p1}%
\end{subfigure}%
\begin{subfigure}{0.33\textwidth}
\includegraphics[width=0.9\linewidth]{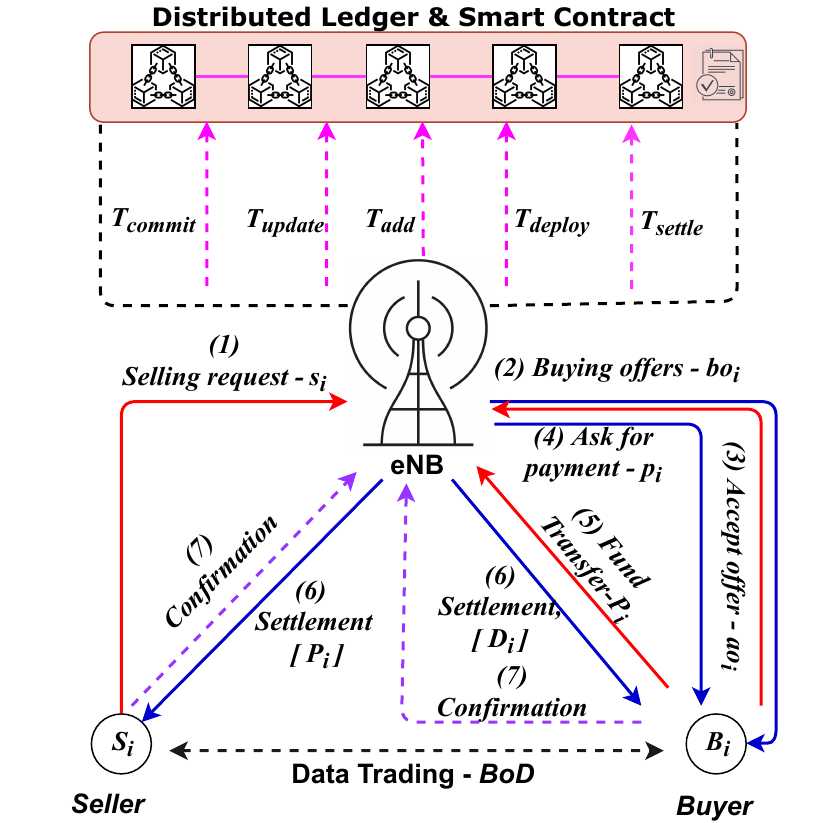}
\caption{\textit{Buying on Demand (BoD) }}
\label{fig:p2}
\end{subfigure}%
\begin{subfigure}{0.33\textwidth}
\includegraphics[width=0.9\linewidth]{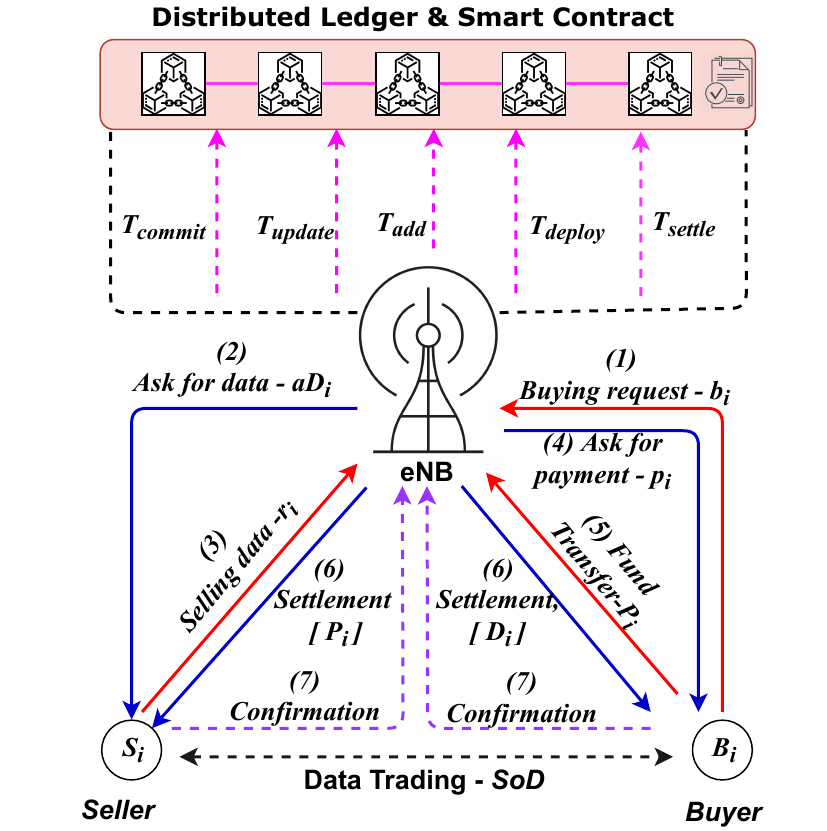}
\caption{\textit{Selling on Demand (SoD) }}
\label{fig:p3}
\end{subfigure}
\caption{Three IoT data trading protocols: (a)\textit{ General Trading protocol (GT), (b)} \textit{Buying on Demand (BoD)}, and (c) \textit{Selling on Demand (SoD)}.}
\label{fig:protocol}
\end{figure*}

\subsection{IoT Data Trading Protocols}
\subsubsection{General Trading (GT)}
The \textit{GT} protocol procedure is shown in Fig. \ref{fig:p1}. In a trade $\mathcal{T}_i$, the buyer $\mathcal{B}_i$ sends a buying request $r_i$ to market via transaction $T_{i, add}$ to express its need in specific data $\mathcal{D}_i$. \rv{After collecting sensing data,} the data producer and seller $\mathcal{S}_i$, \rv{begins} publishing its data information $D_i$\rv{,} to the market. The smart contracts receive the requests from both customers $\mathcal{B}$ and producers $\mathcal{S}$ and then map the buying requests and selling requests to satisfy both parties based on their expected data and price. The buyer $\mathcal{B}_i$ commit\rv{s to} the request with \rv{a} fund transfer via $T_{i, commit}$. After the smart contract receives the payment from $\mathcal{B}_i$, it executes $T_{i, settle}$ to transfer requested $\mathcal{D}_i$ to $\mathcal{B}_i$ and $\mathcal{P}_i$ to $\mathcal{S}_i$. Finally, both $\mathcal{B}_i$ and $\mathcal{S}_i$ confirm to the ledger that they \rv{have} \AL{received} $\mathcal{P}_i$ and $\mathcal{D}_i$, respectively.

A marketplace exchange of streaming IoT data, with a massive amount of data, requests, and a large number of parties, is an appropriate use case for the \textit{GT} protocol. The environmental sensing data such as accurate real-time measurement data for control and alarm systems are exchanged between interested customers.
More specifically, this protocol is used for the aforementioned use case due to its wide range of data continuously being pushed to the market. The open advertisement style of the \textit{GT} protocol is appealing to potential buyers, encouraging the safe buying and selling of IoT data in a decentralized IoT data marketplace.

\subsubsection{Buying on Demand (BoD)}
\textit{BoD} protocol describes a process where the producer $\mathcal{S}_i$ publishes data $\mathcal{D}_i$ to the market for selling via $s_i$ request. The smart contract will broadcast information of received data \rv{from} buying offer $bo_i$ to other parties. For example,  in a buying offer, \PP{$\mathcal{B}_i$ would ask whether \rv{others} are interested in buying $\mathcal{D}_i$.} If \PP{$\mathcal{B}_i$} is interested in $D_i$, it will accept the offer by generat\rv{ing} $T_{i, add}$, and commit by $T_{i, commit}$ when the payment requests from smart contract \rv{is received}. Then, the deal is settled as \textit{GT} protocol via  $T_{i, settle}$. The process of \textit{BoD} protocol is described in Fig. \ref{fig:p2}. 

Vehicle-to-Industry (V2I) emission trading, with a frequent exchange of data between vehicles and the vehicle industry, is an appropriate use case for the \textit{BoD} protocol. In this scenario, the vehicles on the network act as the sellers of their emissions data, e.g., \ch{CO2}, \ch{NOx}, while manufacturers (vehicle industry), \textbf{GoV} e.g air quality management department, and data analytic organizations act as the buyers for maintaining accurate, secure tamper-proof vehicular emissions data.
In V2I, the data being exchanged are used for the purpose of creating a trusted life-cycle emission or fuel economy monitoring.

\subsubsection{Selling on Demand (SoD)}
The \textit{SoD} protocol is described in Fig. \ref{fig:p3}. \AL{In this} case, the smart contract receives the buying requests $b_i$ from a customer \PP{$\mathcal{B}_i$}, but there is no available appropriate data on the ledger to satisfy the requirements from \PP{$\mathcal{B}_i$}. \PP{Hence, the smart contract sends an \emph{ask-for-data}} request $aD_i$ to producers to ask whether they can provide the required data $D_i$. The providers $\mathcal{S}_i$ after a while can gather data from the environment or from other sources then answer to the market by $s_i$ including $D_i$ information as well as price $P_i$. Then, the smart contract asks $\mathcal{B}_i$ for fund transfer with an amount of $P_i$. The $\mathcal{B}_i$ make payment via $T_{i, commit}$. Then $T_{i, settle}$ are executed to complete the deal between $\mathcal{B}_i$ and $\mathcal{S}_i$. Finally, the confirmations are sent to the ledger from both parties. 

\rv{This \textit{SoD} protocol is beneficial, for example, when a party needs a type of data that is not available on the market and there is the need to trade in real time. In the scope of this study, we assume that,  when  a  provider  receives  \emph{ask-for-data} from a smart contract, it can provide the required data to the market. In  real-life  scenarios,  some of the  requests  from customers cannot be satisfied immediately, so these requests are queued in the  systems until  the data is available.}
A Vehicle-to-Vehicle (V2V) use case is appropriate for this protocol where vehicles can purchase traffic information \rv{for a} specific street which drivers expect to \AL{use} in the near future. The vehicles that have the requested information can be traded with the buyers on the road.
Finally, similar to V2I, V2V involves the continuous wireless exchange of IoT data collected from vehicle sensors. The V2V use case contributes in generating a life-cycle emissions or fuel economy monitoring system amongst vehicles. This form of communication helps to manage the safety of the road, as well as increase vehicle awareness.

\begin{figure}[t]
    \centering
    \includegraphics[width = 0.7\linewidth]{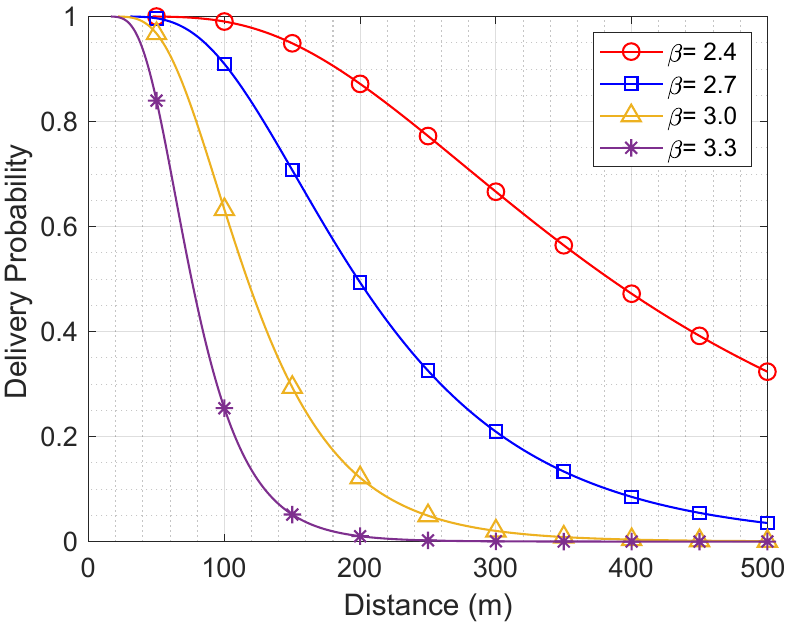}
    \caption{Delivery probability versus distance for a standard deviation $\sigma_{dB}$ = 6 dBs.}
    \label{fig:probd}
\end{figure}

\subsection{Communication System Model} 

We consider an NB-IoT cell with eNB located in its center, including $N$ devices uniformly distributed within the area. 
A data provider or consumer can consist of a single or multiple NB-IoT devices. For simplicity, we assume that each buyer or seller owns a single  NB-IoT device to exchange assets and 
all devices belong to the normal coverage class. The DLT nodes are NB-IoT devices that have more \PP{computational power} than seller/buyer nodes. \LD{In our model, involved sellers and buyers use NB-IoT as wireless network interfaces. In reality, the involved parties can use various wireless interfaces or networks for trading purposes but, our general model and analysis can be applied in these cases because of its modular and versatile design satisfies a broad range of interfaces and networks.}

Our propagation model takes  into account shadowing, but not small-scale fading; which is a sufficient first approximation as detailed physical layer modeling is not the focus of the work. Hence, for a given transmission power $P_t$ and carrier frequency $f$, the received power at a distance $d$ between the base station $BS$ and sensor $i$ is:
\begin{equation}
    P_r(d)= 10 \log_{10}\left[\frac{ P_t G_t G_r\,c^2 }{(4\pi f)^2 d^\beta}\right]+N(0,\sigma_\text{dB})~\text{dB}
\end{equation}
where $G_t$ and $G_r$ are the transmitter and receiver antenna gains, respectively,  $c=3\cdot10^8$~m/s is the speed of light, $N(0,\sigma_\text{dB})$ is a zero-mean Gaussian RV with standard deviation $\sigma_\text{dB}$~dB, and $\beta$ is the path loss exponent. From there, the outage probability at a given distance and receiver sensitivity $\gamma = 3.65\cdot 10^{-10}$ W is: 
\begin{equation}
    p_{out} = 1- Q \left [ \frac{1}{\sigma_\text{dB}} 10 \log_{10} \left( \frac{\gamma (4\pi f)^2 d^\beta}{P_t G_t G_r c^2} \right)  \right]
\end{equation}
Fig. \ref{fig:probd} demonstrates the delivery probability $p_d = 1- p_{out}$ at varying distances, for four different $\beta$ path loss exponent values, a standard deviation of $\sigma_{dB}= 6$ dBs. In this work, we choose $\beta=2.7$ for urban area. \LD{We are aware that the model lacks a mobility aspect, however for this initial work, we have decided to use a simple model as previously described.}

\begin{figure}[t!]
    \centering
    \includegraphics[width=0.8\linewidth]{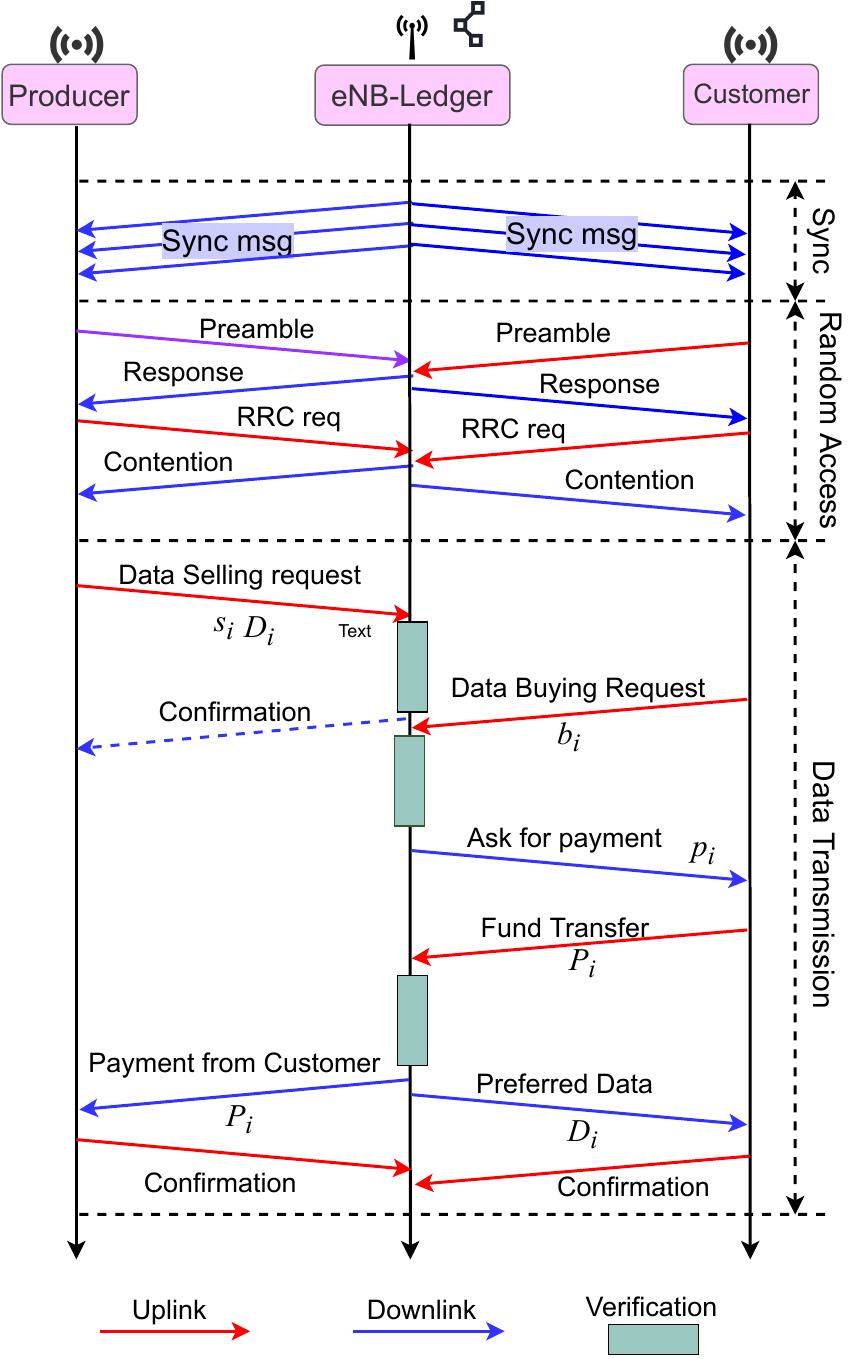}
    \caption{\PP{Communication diagram of the }\textit{GT} protocol.}\vspace{-1em}
    \label{fig:physical}
\end{figure}

The arrival rate of uplink including \PP{selling and buying  requests, respectively,} to the system are: 
$\lambda_s = |\mathcal{S}| T p_s p_d \text{ and } \lambda_b = |{\mathcal{B}| T p_b p_d}$
in which $T$ is number of communication sessions that an IoT device performs daily; $p_s$ and $p_b$ are probability a device request a selling service and a buying service, respectively. When an NB-IoT sensor device attempts to join the network, it first listens for the cell information, e.g, NPSS and NSSS messages to synchronize with the eNB. NB-IoT UEs have only two modes of operation, namely radio-resource control (RRC) idle and RRC connected~\cite{3GPPTS36.331}. In the former, the UEs can only receive the system information from the BS and, only in the latter, data can be transmitted. UEs are in idle mode before initial access to the network, but may also enter this mode during power saving or after an explicit disconnection request. To transition from idle to connected mode, the UEs (clients) must first acquire the basic system information and synchronization as illustrated in the upper part of Fig. \ref{fig:physical}. For this, the UE receives the master information block (MIB-NB) and the system information blocks 1 (SIB1-NB) and 2 (SIB2-NB). These are transmitted periodically through the downlink shared channel (DL-SCH) and carry the basic cell configuration, timing, and access parameters\cite{lsync}. In addition, SIB1-NB carries the scheduling information for the rest of the SIBs. 

After the system information has been acquired, the UEs must perform the RA procedure to transition to RRC connected mode. The RA procedure is a four-message handshake, initiated by the UEs by transmitting a single-tone frequency-hopping pattern, called preamble, through the NB Physical Random Access Channel (NPRACH). In most cases, the RA procedure is contention-based, hence, the preamble is chosen randomly from a predefined pool of up to $48$ orthogonal sub-carrier frequencies. Consequently, the main reasons for an access failure are the lack of power in the transmission and simultaneous transmissions of the same preamble, which leads to collisions. 

After completing the RA procedure, and if the control-plane (CP) cellular IoT (CIoT) is used, UEs may piggyback short UL data packets along with the RRC Connection Setup Complete message. Otherwise, the non-access stratum (NAS) setup must be completed before eNB allocates resources for uplink transmission through the NB Physical UL Shared Channel (NPUSCH) and data can be transmitted. The resource unit (RU) is the basic unit for resource allocation in the NPUSCH and comprises a set of sub-frames in the time domain and sub-carriers in the frequency domain. The downlink (DL) data is transmitted through the NB physical DL shared channel (PDSCH). \PP{Data is exchanged based on the three defined trading protocols, \textit{GT, BoD} and \textit{SoD}. Fig. \ref{fig:physical} shows the physical operations of \textit{GT} protocol as an example.} \LD{
\textit{BoD} and \textit{SoD} could be considered as extensions of the \textit{GT} protocol, those protocols are especially beneficial when the data is not available in the market.}

\subsection{Performance metrics}

\PP{Latency and the time required to complete a trade is one of the most important concerns of involved users. Latency directly influences the amount of time it takes for a trader to interact with the data market, the timely 
reception of relevant market information and the ability to act upon its receipt. The spread of the automatized data trading amplifies the impact of latency in terms of its competitive advantage. On top of this, IoT environments should be characterized with high energy efficiency. All these factors have motivated this investigation on the   total E2E latency and energy consumption to complete a trade $\mathcal{T}_i$.} 

\subsubsection{Latency} The latency to complete a trade $\mathcal{T}_i$ between seller and buyer are formulated as:
\begin{equation} 
    L_{\mathcal{T}_i}   = L_{UD} +  L_{DLT},
\end{equation}
where $L_{UD}$ is the transmission latency between $\mathcal{S}_i$ and $\mathcal{B}_i$ which act as light nodes and full DLT nodes; While, $L_{DLT}$ represents the DLT mining and synchronization latency. In detail, $L_{UD} = L^u + L^d$, where $L^u$, $L^d$ are NB-IoT uplink and downlink latency, respectively; $L_{DLT} = L_{v} + L_{DLTsync}$, where $L_{v}$ is block verification time at DLT nodes, and $L_{DLTsync}$ is synchronization time between DLT nodes via NB-IoT connectivity.

\subsubsection{Total energy consumption} Similarly, the energy consumption model of a trade includes the energy consumption for uplink $E^u$, downlink $E^d$ transmission between NB-IoT sensors with DLT full nodes, among DLT full nodes, and the energy consumed in verification process known as mining in DLT nodes $E_{DLT}$.
\begin{equation} 
    E_{\mathcal{T}_i}   = E_{UD} +  E_{DLT}
\end{equation}
where $E_{UD}$ and $E_{DLT}$ are energy consumed by communication between sellers/buyers and DLT nodes and the energy performed among full DLT nodes, respectively. \PP{The transmission power and latency depend significantly on the physical deployment, such that we analyze  both analyze the resource consumed in physical communication and the application layer.} In next parts, we formulate the latency and energy consumption of each process.
\section{NB-IoT Transmission Latency and Energy Consumption Models}

As described in the previous section, the total E2E latency includes two parts, the latency of transmissions of uplink and downlink between buyers/sellers and DLT nodes, \rv{where} latency occurs in the DLT verification process. For the first part, we define an adapted queuing model for DLT-based NB-IoT\rv{,} based on \rv{the} queuing model of the NB-IoT access network \cite{energymodel}, the uplink and downlink radio resources are modeled as two servers which visit and serve their traffic queues in both directions. 

\subsubsection{End-to-End latency} The E2E latency of NB-IoT uplink and downlink can be formulated as: 
\begin{align}
    \begin{split}
        L_{UeD}  &= L^{u} + L^{d} \\&= L^{u}_{sync} + L^{u}_{rr} + L^{u}_{tx}  + L^{d}_{sync} +  L^{d}_{rr} + L^{d}_{rx}
    \end{split}
\end{align}
where $L^u_{sync}$, $L^u_{rr}$, $L^u_{tx}$, $L^d_{synch}$,$L^d_{rr}$, $L^d_{rx}$ are energy consumption of synchronization, resource reservation, and data transmission of uplink and downlink, respectively. $L^u_{sync}$ has been defined in \cite{lsync} with the values of $0.33s$. $L_{rr}$ is given as:%
\begin{equation}
    L_{rr} = \sum_{l=1}^{N_{r_{max}}} (1-P_{rr})^{l-1} P_{rr}l(L_{ra} + L_{rar})
\end{equation}
in which $N_{r_{max}}$ is the maximum number of attempts, $P_{rr}$ is the probability of successful resource reservation in an attempt, $L_{ra} = 0.5t + \tau$, is the expected latency in sending an RA control message, $\tau$ is the unit length and equal to the NPRACH period for the coverage class 1 which is varied from $40$\,ms to $2.56$\,s \cite{lsync}, and $L_{rar}=0.5d + 0.5 \mathcal{Q} fu +u $, is the expected latency in receiving the RAR message, where $\mathcal{Q}$ are requests waiting to be served. 

In the following, we provide a simple technique based on \emph{drift approximation}~\cite{drift_approx} to calculate $P_{rr}$ recursively. Therefore, we treat the mean of the random variables involved in the process as constants. Besides, we assume that sufficient resources are available in the PDCCH so that failures only occur due to collisions in the PRACH or to link outages.

Let $\lambda^a=\lambda^u+\lambda^d$ be the arrival rate of access requests per PRACH period and $\lambda^a(l)$ be the mean number of devices participating in the contention with their $l$-th attempt. Note that in a steady state $\lambda^a(l)$ remains constant for all PRACH periods. Next, let 
$\lambda^a_{tot}=\sum_{l=1}^{N_{r_{max}}} \lambda^a(l)$
and that the collision probability in the PRACH can be calculated using the drift approximation for a given value of $\lambda^a_{tot}$ and for a given number of available preambles $K$ as:
\begin{equation}
    P_\text{collision}(\lambda^a_{tot})=1-\left(1-\frac{1}{K}\right)^{\lambda^a_{tot}-1}\approx 1-e^{-\frac{\lambda^a_{tot}}{K}}.
\end{equation}
\rv{From there, we approximate the probability of resource reservation as a function of $\lambda^a_{tot}$ as
$P_{rr}(\lambda^a_{tot})\approx p_d\,    e^{-\frac{\lambda^a_{tot}}{K}}.$ }
This allows us to define $\lambda^a_{tot}$ as: 
\begin{equation}
\lambda^a_{tot}= \lambda^a +\left(1-P_{rr}(\lambda^a_{tot})\right)\sum_{l=2}^{N_{r_{max}}} \lambda^a(l)
\label{eq:lambdatot}
\end{equation}%
since $\lambda^a(l)=\left(1-P_{rr}(\lambda^a_{tot})\right)\lambda^a(l-1)$ for $l\geq2$ and $\lambda^a(1)=\lambda^a$. Finally, from the initial conditions $\lambda^a(l)=0$ for $l\geq2$, the values of $\lambda^a(l)$ and $\lambda^a_{tot}$ can be calculated recursively by: 1) applying~\eqref{eq:lambdatot}; 2) calculating $P_{rr}(\lambda^a_{tot})$ for the new value of $\lambda^a_{tot}$; and 3) updating the values of $\lambda^a(l)$. This process is repeated until the values of the variables converge to a constant value. The final value of $P_{rr}(\lambda^a_{tot})$ is simply denoted as $P_{rr}$ and used throughout the rest of the paper.

Assuming that the transmission time for the uplink transactions follows a general distribution with the first two moments $l_1, l_2$, then first two moments of the distribution of the packet transmission time are $s_1 = \left(f_1l_1\right)/\left(\mathcal{R}w\right)$, and  $s_2 = \left(f_1l_2\right)/\left(\mathcal{R}^2w^2\right)$. Applying the results from \cite{ltx}, considering $L_{tx}$ as a function of scheduling of NPUSCH, we have: 
\begin{equation}
    L_{tx} = \frac{f\lambda^us_1s_2}{2s_1(1-fGs_1)} + \frac{f\lambda^us_1^2}{2(1-f\lambda^us_1)} + \frac{l_1}{\mathcal{R}^uw}
\end{equation}
where $\mathcal{R}^u$ is the average uplink transmission rate, $\lambda^u = \lambda_s +\lambda_b$, and $f(\lambda_s + \lambda_b)s_1$ is the mean batch-size. The latency of data reception is defined as: 
\begin{equation}
    L_{rx} = \frac{0.5Fh_1t^{-1}}{h_1(1-Fht^{-1})} + \frac{Fh_1}{1-Fht^{-1}} + \frac{m_2}{\mathcal{R}^dy}
\end{equation}
in which, $h_1=fm_1(\mathcal{R}^dy)^{-1}$, $h_2 = fh_2^2m_2 ((\mathcal{R}^d)^2y^2)^{-1}$ are two moments of distribution of the packet transmission time, assuming that Assuming that packet length follows a general distribution with moments $m_1$, $m_2$, $F=f\lambda^d t$, $\mathcal{R}^d$ is downlink data transmission rate. 

\subsubsection{Energy consumption} The energy consumption of the protocol 1 are formulated as below:
\ilm{\begin{align}
    \begin{split}
        E_{UD}   =  E^{u} + E^{d} =& E^{u}_{sync} + E^{u}_{rr} + E^{u}_{tx} +E^{u}_{s} \\ & + E^{d}_{sync} +  E^{d}_{rr}  + E^{d}_{rx} + E^{d}_{s}
    \end{split}
\end{align}}%
In which $E^u_{sync}$,$E^u_{rr}$, $E^u_{rr}$, $E^d_{sync}$,$E^d_{rr}$, $E^d_{rx}$ are energy consumption of synchronization, resource reservation, and data transmission of uplink and downlink, respectively. We have:
\begin{align}
 & E_{sync} = P_l \cdot L_{sync} \\
 & E_{rar} = P_l \cdot L_{rar} \\
 &  E_{rr} = \sum_{l=1}^{N_{max}} (1-P_{rr})^{l-1}  \cdot P_{rr}  \cdot (E_{ra} + E_{rar}) \\
 & E_{ra} = (L_{ra} - \tau) \cdot P_I + \tau \cdot (P_c + P_e P_t) \\
 & E_{tx} = (L_{tx} - \frac{l_a}{\mathcal{R}^uw}) \cdot P_I + (P_c + P_e P_t)\frac{l_a}{\mathcal{R}^uw} \\
 & E_{rx} = (L_{rx} - \frac{m_1}{\mathcal{R}^dy}) \cdot P_I + P_l \frac{m_1}{\mathcal{R}^dy}
\end{align}
in which, $P_e$, $P_I$, $P_c$, $P_l$, and $P_t$ are the power amplifier efficiency, idle power consumption, circuit power consumption of transmission, listening power consumption, and transmit power consumption, respectively. %
\begin{figure}[t]
    \centering
    \includegraphics[width=0.8\linewidth]{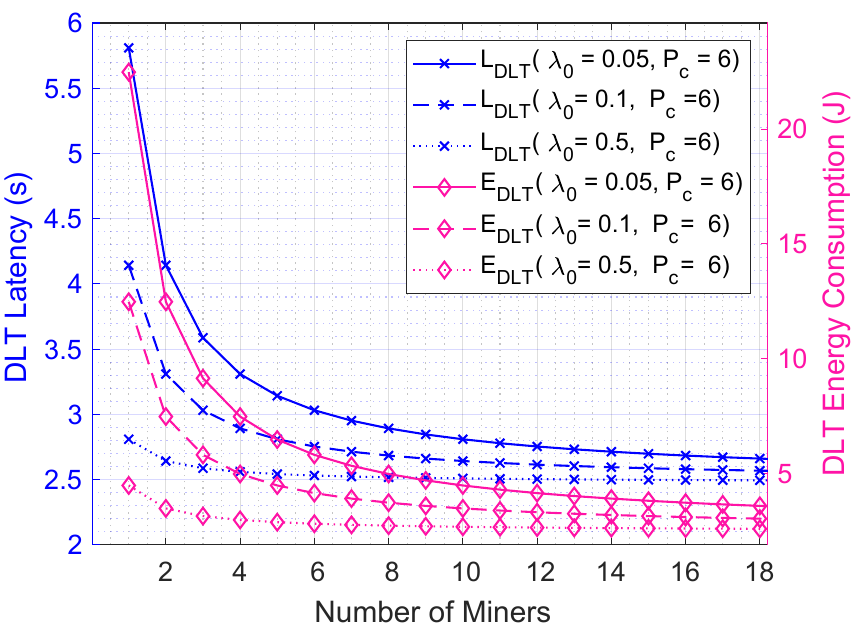}
    \caption{DLT performance in latency and energy consumption}
    \label{fig:dltperformance}
\end{figure}%

\section{Resource consumption model of DLT verification process in NB-IoT environment}
Consider a DLT network that includes $M$ miners. These miners start their Proof-of-work (PoW) computation at the same time and keep executing the PoW process until one of the miners completes the computational task by finding the desired hash value \cite{bitcoin}. When a miner executes the computational task for the POW of current block, the time period required to complete this PoW can be formulated as an exponential random variable $W$ whose distribution is $f_W(w) = \lambda_c e^{-\lambda_c w}$, in which $\lambda_c= \lambda_0 P_c$ \rv{represents} the computing speed of a miner, $P_c$ is power consumption for computation of a miner, and $\lambda_0$ is a constant scaling factor. Once a miner completes its PoW, it will broadcast messages to other miners, so that other miners can stop their PoW and synchronize the new block. %
\begin{equation}
    L_{tM} = L_{newB} + L_{getB} + L_{transB}
\end{equation}

\rv{In (18),} $L_{newB}$, $L_{getB}$, and $L_{transB}$, are latencies of sending hash of new mined block, requesting new block from neighboring nodes, and new block transmission, respectively. $L_{newB}$ and $L_{transB}$ are computed using uplink transmission, while $L_{getB}$ is computed based on downlink transmission as described in previous section. 

For the PoW computation, a miner $i*$, first finds out the desired PoW hash value, $i*= \min_{i \in M } w_i$. The fastest PoW computation among miners is $W_{i*}$, the complementary cumulative probability distribution of $W_{i*}$ could be computed as $Pr(W_{i*} > x)= Pr(\min_{i \in M} (W_i) > x) = \prod_{i=1}^H Pr(W_i > x) = (1 - Pr(W < x))^M$. Hence, the average computational latency of miner $i*$ is described as: 
\ilm{
\begin{IEEEeqnarray}{rCl}
    L_{W_{i*}} &=& \int_0^\infty (1 - Pr(W \leq x))^M \dd x \\
    &=& \int_0^\infty e^{-\lambda_cMx}\dd x = \frac{1}{\lambda_cM}
\end{IEEEeqnarray}}

The total latency required from DLT verification process is $L_{DLT} = L_{tm} + L_{W_{i*}}$. 
The average energy consumption of DLT to finish a single PoW round is:
\begin{equation}
    E_{DLT} = P_c L_{W_{i*}} + P_{t} L_{tm}
\end{equation}
The performance of DLT system is shown in Fig. \ref{fig:dltperformance}. The figure demonstrates that the energy consumed and latency by DLT nodes are reduced with the number of miners. \rv{Contrarily}, as the number of miners \rv{increase}, this leads to a higher probability that miners verify transactions, and the mining speeds \rv{increase} as well. 


\subsection{Analysis of data trading protocols}
In this section, the E2E latency and energy consumption of three protocols are formulated and compared approximately. The resource consumed by each data trading protocol is separated into two parts, namely, 1) the connectivity between $\mathcal{S}_i$ and $\mathcal{B}_i$ acting as light nodes in DLT network with full nodes, and 2) the communication among DLT full nodes.

The E2E latency of trade $\mathcal{T}_i$ using \textit{GT} protocol including the transmission latency between $\mathcal{B}_i$, $\mathcal{S}_i$ and DLT verification nodes is described as below: %
\begin{equation}
    L^{P1}_{\mathcal{T}_i}  = L^{P1}_{UD}+ L^{ P1}_{DLT} =  L^{u,P1} + L^{d,P1} + L^{ P1}_{DLT} 
\end{equation}%
\begin{equation}
    E^{P1}_{\mathcal{T}_i}   =  E^{P1}_{UD}+ E^{ P1}_{DLT} = E^{u,P1} + E^{d,P1} + E^{P1}_{DLT}
\end{equation}%
Assuming that $L^{u,P1}_{sync} = L^{d,P1}_{sync} =$ 0.33 s, $L^{u,P1}_{rr}$ and $L^{d,P1}_{rr}$ are computed as (7), $L^{u,P1}_{tx}$ and $L^{d,P1}_{rx}$ are calculated based on (8) and (9) with the defined packet length of uplink and downlink. 

Then, the battery lifetime of an NB-IoT device can be computed as below: 
 $   BTL = E_0 \left[Tp^u(E^u)  + Tp^dE^d \right]^{-1}, $
where $E_0$ is the energy storage on the device battery. Similarly, the performance of \textit{BoD} protocol and \textit{SoD} protocol can be formulated as \textit{GT} protocol. 

\section{Performance Evaluation}
In this section, we will introduce the settings in terms of simulations and experiments. Then, we analyze the performance of proposed trading protocols in terms of latency and battery lifetime. 

\subsection{Experiment Settings}
In this section, we evaluate the derived data trading model, compare and analyze the designed trading protocols. In order to evaluate the derived model and compare the three proposed protocols, we setup a network with $N=10000$ NB-IoT devices, where devices randomly play roles as sellers or buyers. We validate the results via Monte Carlo Simulations, where we run 1000 realizations for each trading protocol and \rv{experiment}. The buyer nodes and seller nodes generate requests following a Poisson distribution process with rates of $\lambda_b$ and $\lambda_s$, respectively. The number of buying and selling $\mathcal{T}_i$ requests per day varied from 1 to 20, $\mathcal{T}_i = [1,20]$. \rv{Additionally, the number of buyer and seller nodes varied and remained less than $N$.} The transmission power \rv{in the experiments are denoted} as $P_t=0.2$W, $E_0=1000$. The number of DLT miners are up to 20 miners \rv{at} maximum, $M=[1,20]$. 

\begin{table}[t]
\centering
\caption{Comparison in Smart Contract Execution Cost}
\begin{tabular}{@{}c c c c c@{}} 
\toprule
 \textbf{Protocols} & \textbf{Operations} & \textbf{Ether\ilm{$\cdot10^{-4}$}} & \textbf{Gas Cost} & \textbf{Approx. USD} \\ [0.3ex] 
\midrule
\textit{GT}& Deploy SC &  \ilm{$\approx$1.2} & 1132443 & 0.2862 \\ 
\midrule 
\textit{BoD} & Deploy SC &  \ilm{$\approx$1.3} & 1268369 & 0.2879 \\ 
\midrule 
\textit{SoD} & Deploy SC &  \ilm{$\approx$1.4} & 1582349 & 0.3783 \\ 
\bottomrule
\end{tabular}
\begin{tablenotes}
      \small
      \item * 1 Ether = $10^9$ Gwei; 1 USD = 4,182,471.9949 Gwei
    \end{tablenotes}
    \label{tab:costs}
\end{table}

\begin{figure}[t]
\centering
\begin{subfigure}{0.5\textwidth}
\centering
    \includegraphics[width=0.9\linewidth]{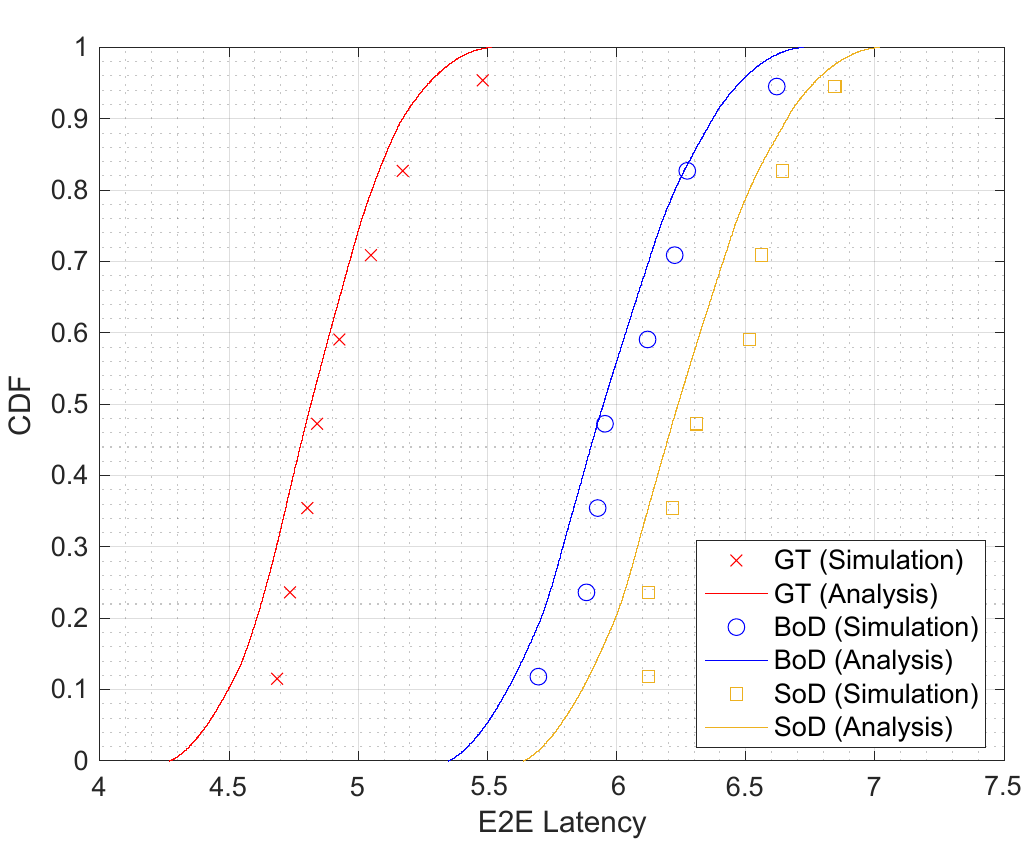} 
    \caption{M = 20 }   
    \label{fig:e2e1}%
\end{subfigure}

\begin{subfigure}{0.5\textwidth}
\centering
    \includegraphics[width=0.9\linewidth]{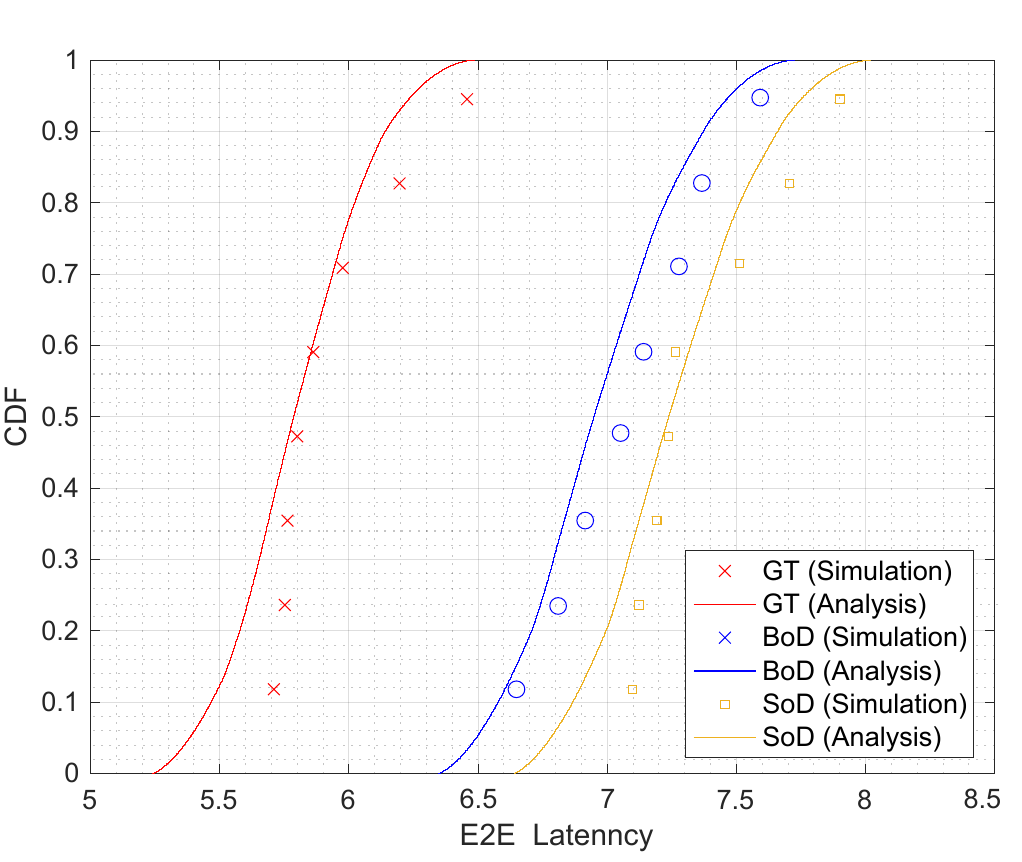}
    \caption{M = 5.}
    \label{fig:e2e2}
\end{subfigure}
    \caption{Impact of number of DLT miners to latency of trading strategies.}
    \label{fig:e2e}
\end{figure}

\subsection{Cost of Smart Contracts}
The proof of concept for the three proposed trading protocols are deployed in Ganache\footnote{https://www.trufflesuite.com/ganache} Ethereum network to evaluate the complexity and the cost of execution of different trading strategies. The smart contracts are implemented and deployed using Remix IDE\footnote{https://remix.ethereum.org/}. \rv{In the Ethereum platform, any operation or transaction execution that changes the Blockchain or its state requires that the involved parties pay a fee called \emph{gas}. The gas terminology in Ethereum charges the execution of every operation to guarantee that smart contracts running in Ethereum Virtual Machine (EVM)\cite{ethereum} will be eventually terminated. These costs are calculated by using the amount of gas executed and the unit of gas price. The gas required during any activity reflects the computational complexity or size of the smart contracts, while the gas prices are determined by the Ethereum miners in the network. Each operation or execution on the EVM charges a certain amount of gas and not all transactions are created cost equally.} In this work, we used Gwei\footnote{https://www.cryps.info/} to evaluate the cost of different operations in the trading process. 

Table\,\ref{tab:costs} shows the cost of the three protocol deployment and transaction costs to complete a deal between a seller and a buyer. We observe that the approximate cost in USD for \textit{GT} is the cheapest in comparison to \textit{BoD} and \textit{SoD} protocols. The cost of smart contract execution is generally expensive, therefore, it is preferred to use the \textit{GT} protocol. In an environment with a massive number of involved parties and transactions (e.g, marketplace), the transactions are executed autonomously to reduce costs using the available resources. While, \textit{BoD} and \textit{SoD} are preferred when the users have requests with specific resources. %

\begin{figure}[t]
    \centering
    \includegraphics[width=0.9\linewidth]{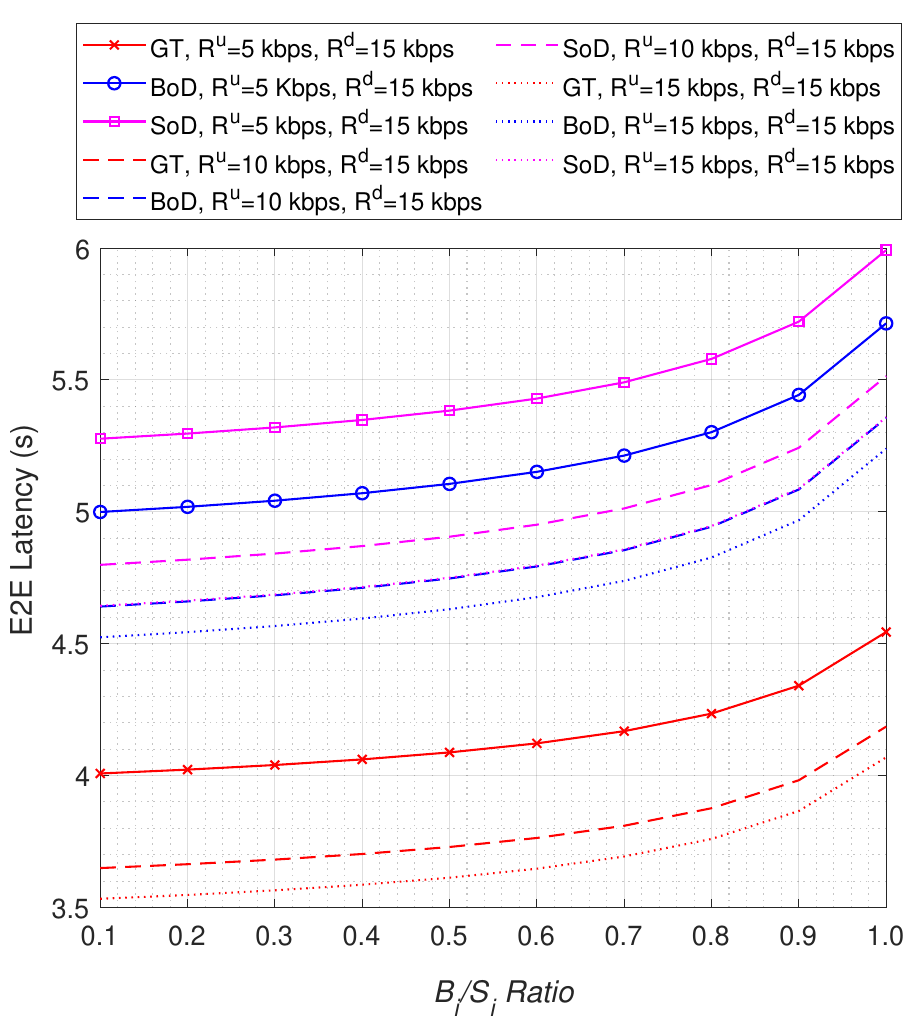}
    \caption{Impact of $B_i / S_i$ ratio }
    \label{fig:bsratio}
\end{figure}

\subsection{Latency to complete a deal}

\subsubsection{Impact of number of Miners} Fig. \ref{fig:e2e} shows the latency of three trading protocols. Both the analysis and the simulation results show that the \textit{SoD} protocol has higher latency to complete a deal between $S_i$ and $B_i$ because of extra steps. Note that the comparison is evaluated approximately because the latencies \rv{depend} on various factors such as the number of DLT miners, the length of blocks, and level of difficulty. The verification latency of DLT miners is measured based on the Ganache Ethereum network. In \textit{GT} protocol, the smart contracts map selling requests $r_i$ with available $D_i$ stored in the ledger and make a deal between $S_i$ and $B_i$ immediately, so that it guarantees efficient trading in the market. The average latency to complete a deal of \textit{GT} protocol is around $4.5$ seconds including latency of NB-IoT and DLT procedures. The \textit{BoD} and \textit{SoD} latencies are higher because of extra procedures necessary to gather required information between customers and producers. We observe that \textit{GT} could be used in terms of applications which require low latency. The downside of \textit{GT} protocol is that the data requests \rv{must} always be available to settle the trade, so that it is matched with applications (e.g smart metering) where the type of information is fixed.

\subsubsection{Seller and Buyer Ratio} The impact of ratio between the number of sellers and buyers are demonstrated in Fig. \ref{fig:bsratio}. The figure also shows a comparison between trading protocols under varying NB-IoT uplink transmission rate, $R^u= \{5, 10, 15\}$ Kbps and fixed downlink data rate at $R^d=15$ Kbps. The results show that i) the increase in the number of buyers requires more delay to complete a trade, and ii) in contracts, increasing data rates help to provide a faster service.

\begin{figure}[t]
    \centering
    \includegraphics[width=0.9\linewidth]{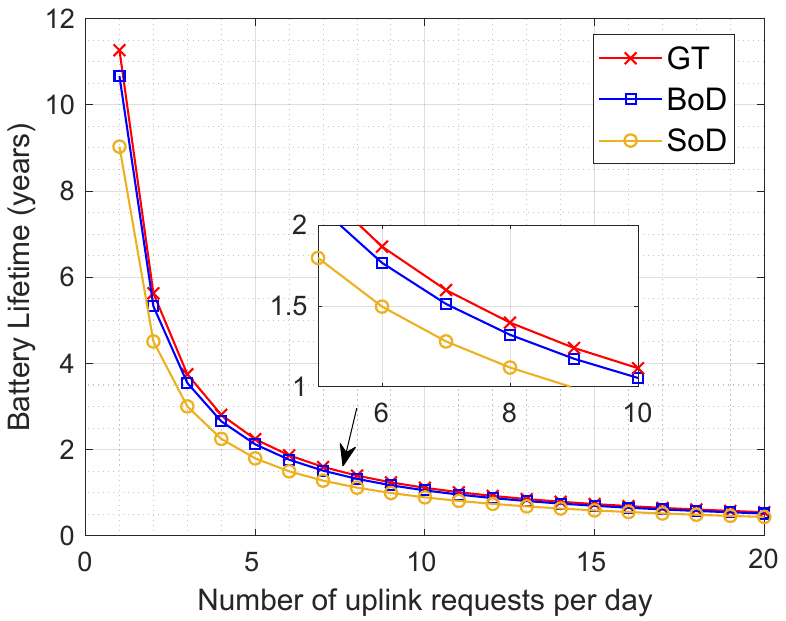}
    \caption{NB-IoT Battery lifetime.}
    \label{fig:btl}
\end{figure}

\subsection{Battery lifetime of NB-IoT devices}
In general, the power consumption of battery lifetime during a reporting period depends on length of data transmitting, bandwidth, MCL, latency, and RF module. Hence, the power consumption of one trading protocol will be higher or lower than the other depending on the values of these parameters. The battery lifetime capabilities of NB-IoT devices among three trading protocols are compared and demonstrated in Fig. \ref{fig:btl}. The number of uplink requests are varied from 1 to 20 requests per day. We observe that the number of requests per day significantly impacts to the battery lifetime of NB-IoT devices. \rv{In fact,} the battery lifetime of around 10 years can be \rv{achieved} with one report per day, however, for more frequent transmissions (e.g. 8 requests per day) the battery lifetime is reduced to around 1 year. Specifically, the \textit{GT} trading protocol achieves over 11 years for 1 report per day, while \textit{BoD} and \textit{SoD} achieve around 10 years and 9 years, respectively. The fact is that applications such as smart metering, smart parking using NB-IoT connectivity do not require frequent updates from sensors. In terms of increasing number of requests daily up to 5, the battery lifetime is reduced significantly to around 2 years. Because for each buying or selling request, the NB-IoT devices start running protocol with multiple operation until the trade is settled. 

\section{Conclusion}
\PP{In this paper, we proposed the first benchmarking framework for evaluating data trading protocols. The framework includes a model and analysis of systematic DLT-based IoT data smart trading protocols in massive NB-IoT deployments. We \rv{have} proposed and analyzed three IoT data trading protocols named \textit{General Trading}, \textit{Buying on Demand}, and \textit{Selling on Demand}. Considered collectively, these protocols cover a wide range of interesting scenarios, such as carbon emission trading or monitoring of vehicle emissions. We have conducted a comprehensive analysis of these protocols in terms of communication and evaluated end-to-end latency, battery lifetime, and resource consumption.}
\PP{In terms of performance, each protocol is tailored to a different scenario.}  \LD{We conclude that the \textit{GT} protocol should be used as primary protocol in a data marketplace where massive amounts of data are available. Additionally, the \textit{BoD} and \textit{SoD} protocols can be interchangeably used when there are particular demands from either buyers or sellers}. 

To the best of our knowledge, this is the first work of its kind, providing a general benchmark framework for data trading protocols in IoT environments. \AL{In the next iteration of this work}, we will first \PP{consider} more elaborate utility models for the parties involved in trading. \LD{Second, we will evaluate the performance of trading schemes in diverse network interfaces and real-life networks.}

\bibliographystyle{ieeetr}
\bibliography{revised-Manu}

\end{document}